\documentclass[10pt]{article}
\usepackage{amsfonts,latexsym,color,array}

\newcommand{\R}{\mathbb R}

\newcounter{shownewstuffflag}

\newcommand{\startnewstuff}{\ifnum\value{shownewstuffflag}>0\color{blue}\fi}
\newcommand{\finishnewstuff}{\ifnum\value{shownewstuffflag}>0\color{black}\fi}

\newcounter{oldeq}

\def\beq{\begin{equation}}
\def\eeq{\end{equation}}

\newtheorem{theorem}{Theorem}
\newtheorem{lemma}{Lemma}
\newtheorem{proposition}{Proposition}

\begin{document}
\title{Topologically General $U(1)$ Symmetric Vacuum Spacetimes with AVTD Behavior}

\author{Yvonne Choquet-Bruhat \\ Universit\'e de Paris \and 
James Isenberg\\ University of Oregon  \and
Vincent Moncrief
\\ Yale University}

\date{ }

\maketitle

\begin{abstract} We use Fuchsian methods to show that, for any two dimensional manifold $\Sigma^2$, there is a large family of  $U(1)$ symmetric solutions of the vacuum Einstein equations on the manifold $\Sigma \times S^1 \times \mathbb{R}$, each of which has AVTD behavior in the neighborhood of its singularity.

\end{abstract}

\section{Introduction}
\label{se:intro}

In \cite{IM02}, Isenberg and Moncrief obtain a family of $U(1)$ symmetric cosmological solutions of the vacuum Einstein equations which have asymptotic velocity term dominated (``AVTD") behavior in the neighborhood of their singularities. Such behavior, which has been studied in a number of classes of spacetimes (See, for example \cite{KR98}, \cite{AR01}, and \cite{DHRW02}.), has been found to be useful for the study of the strong cosmic censorship conjecture and similar questions \cite{IM92}. Note, however, that little is known about the long time evolution of these spacetimes away from the singularity; indeed, the proof \cite{IM02} that there are $U(1)$ symmetric solutions with AVTD behavior relies on the use of Fuchsian methods, which only guarantee existence of the solutions in a neighborhood of the singular region.

In \cite{CBM03} \cite{CB04}, Choquet-Bruhat and Moncrief obtain a family of $U(1)$ symmetric cosmological solutions of the vacuum Einstein equations which expand for an infinite amount of proper time, and which asymptotically approach (in the expanding direction) a flat spacetime. In this case, little is known about the behavior of these spacetimes in the contracting direction; a singularity generally forms, but it is not known whether the gravitational fields show AVTD behavior in the neighborhood of the singularity.

Are there solutions which combine these two behaviors: AVTD behavior near the singularity in the contracting direction, and global existence with asymptotic approach to a flat spacetime in the expanding direction? If one attempts to show that such solutions exist by somehow combining the results of \cite{CBM03} with those of \cite{IM02} cited above, one immediately runs into a problem: The AVTD results    of  \cite{IM02} are formulated for spacetimes that are topologically $T^3\times \R$, while the global existence and stability results of \cite{CBM03} work with spacetimes that are topologically $S^1\times \Sigma^2 \times \R$, where $\Sigma^2$ is a surface of genus greater than one. The question then becomes whether either of these results can be adapted to alternative spacetime topologies. In this work we show that for the AVTD results this is indeed the case. 

We show, using a global in space formulation which is independent of the topology, that for any surface $\Sigma^2$, there is a family of $U(1)$ symmetric vacuum spacetimes on  $S^1\times \Sigma^2 \times \R$ which have AVTD behavior in a neighborhood of their singularities. To obtain this result, we have had to add a restriction beyond the polarization (or half-polarization) condition which is imposed in \cite{IM02}. This restriction, which we spell out in section 4, controls the asymptotic (in time) conformal class of the spatial geometry transverse to the orbits of the $U(1)$ isometry group. As a result, one can study (globally in space)  the dynamics of the geometry near the singularity, regardless of the topology of $\Sigma^2.$

The proof of our results here, like that used to prove the main results in \cite{IM02}, relies on Fuchsian methods. We carry out the proof in sections 4 and 5, after describing the spacetimes of interest in section 2, and discussing the velocity term dominated equations and solutions for these spacetimes in section 3. We state our main theorem in section 5, and then make concluding remarks in section 6 regarding the work that needs to be done if we hope to find solutions with both AVTD behavior in the contracting direction and global existence in the expanding direction.

\section{Polarized $U(1)$ Symmetric Vacuum Spacetimes }
\label{se:sptms}

It has long been known that the vacuum Einstein equation system on a spacetime $(^4M, ^4g)$ with a $U(1)$ isometry group acting spatially is equivalent to a coupled system of the Einstein equations and a wave map on the reduced spacetime $(^3M, ^3g)$. Further, if we assume that the spacetime $(^4M, ^4g)$ is \textit{polarized} in the sense that the  3-planes orthogonal to the orbits of the isometry group $U(1)$ are surface-forming, then it follows that the wave map reduces to a (real) scalar field satisfying the wave equation on $(^3M, ^3g)$.

To be more explicit, we choose a smooth surface $\Sigma^2$ (of arbitrary genus), we set $^4M=S^1 \times \Sigma^2 \times \R$,  we choose $\theta$ as the coordinate for the circle (the orbits of the $U(1)$ action), and we write the metric $^4g$ in the form 
\begin{equation}
^4g= e^{-2\phi } (^{3} g) +e^{2\phi }d\theta ^{2},
\end{equation}
where $^{3} g$ is a Lorentz signature metric on $\Sigma^2 \times \R$, and $\phi$ is a scalar field on the same (three dimensional) spacetime manifold (both independent of $\theta$).  The vacuum Einstein equations on $^4M$ now reduce to the system 
\begin{equation} 
\label{2+1Einsteq}
R_{\alpha \beta} [^3 g]=2 \partial_\alpha \phi\partial_\beta \phi
\end{equation}
\begin{equation}
\label{2+1waveq}
D\cdot D \phi=0,
\end{equation}
where $D$ is the (three dimensional) spacetime covariant derivative corresponding to $^3 g$,  $D\cdot D$ is the corresponding D'Alembertian (wave) operator, and the Greek indices $\alpha, \beta$, etc, take the values $\{0,1,2\}$. 

We wish to reformulate  the Einstein system (\ref{2+1Einsteq})-({\ref{2+1waveq}) in $2+1$ dynamical form. We first rewrite the $2+1$ spacetime metric in the form
\begin{equation}
\label{2+1metric}
^3 g= -N^2 dt^2 + \gamma,
\end{equation}
where $N$ denotes the lapse, $\gamma$ is a (time-dependent) Riemannian metric on $\Sigma^2$, and we have (without loss of generality) assumed that the shift is zero. The dynamical variables now consist of the metric $\gamma$ together with the extrinsic curvature
\begin{equation}
k_{ab} =-\frac{1}{2N} \partial_t \gamma_{ab}
\end{equation}
(Here and below the Latin indices $a,b$, etc, take the values $\{1,2\}$;  they denote the indices for a coordinate basis on the 2-surfaces $\Sigma^2$.), 
and the scalar field $\phi$ together with 
\beq
p=\frac{1}{N} \partial_t \phi.
\eeq
 The lapse is not a dynamical variable; its evolution may be freely chosen, and corresponds to a gauge or coordinate choice.
 
It is useful for later purposes to replace the variables $(N, \gamma)$ by the equivalent set $(\lambda, \sigma)$, with the transformation between these two sets of variables given by 
\begin{equation}
\label{Nlambda}
N=e^{\lambda}
\end{equation}
and 
\begin{equation}
\label{gammasigma}
\gamma_{ab}=e^{\lambda}\sigma_{ab}.
\end{equation}
We note that this replacement of $(N, \gamma)$ by $(\lambda, \sigma)$ does not imply any restrictions on the fields, and does not entail any choice of gauge. We also note that in terms of these new variables, we have 
\beq
\label{kdef}
k_{ab} = -\frac{1}{2}(\partial_t \sigma_{ab} + \sigma_{ab} \partial_t \lambda),
\eeq
\beq
\tau = \gamma^{ab}k_{ab} =-e^{-\lambda}(\partial_t \lambda + \frac{1}{2}\psi),
\eeq
where for convenience we use the definition $\psi :=\sigma^{ab}\partial_t\sigma_{ab}$,  
and
\beq
p=e^{-\lambda}\partial_t \phi.
\eeq

As usual for the $n+1$ dynamical formulation of Einstein's theory (here, $n=2$), the field equations split into constraints and evolution equations. The three constraint equations for this reduced system take the form 
\begin{equation}
\label{constr1}
R(\gamma)-k^{ab}k_{ab}+\tau^2=2p^2 +2\gamma^{ab}\partial_a \phi \partial_b \phi
\end{equation}
and 
\beq
\label{constr2}
\nabla_{b}k_{a}^{b}-\partial _{a}\tau =-2p \partial _{a}\phi,
\eeq
while the evolution equations can be written as 
\beq
\label{evolnk}
\partial_t k_{ab} =NR_{ab}-2Nk_{ac}k^c_b +N \tau k_{ab}-\nabla_a\partial_bN-2N\partial_a\phi\partial_b\phi
\eeq
\beq
\label{evolnp}
\partial_t p= -\frac{1}{2} p \psi  - p \partial_t \lambda+ e^{\lambda} \gamma^{ab}( \nabla_a \nabla_b \phi + \nabla_b \phi \partial_a \lambda),
\eeq
together with 
\beq
\label{evolngamma}
\partial_t \gamma_{ab}= -2e^{\lambda} k_{ab}
\eeq 
and 
\beq
\label{evolnphi}
\partial_t \phi= e^{\lambda} p.
\eeq
Here $\nabla$ denotes the covariant derivative compatible with the metric $\gamma$, $R_{ab}$ is the corresponding Ricci curvature, and $R:=\gamma^{ab}R_{ab}$ is the corresponding scalar curvature.  

We have not yet chosen a gauge condition; we do that below. We note, however $g^{\alpha \beta} \Gamma^t_{\alpha \beta}  = \frac{1}{2} \psi e^{-2 \lambda}$, so the system is in time harmonic gauge if and only if  $\psi=0$.

As noted in the Introduction, in this paper we impose an extra restriction on the class of $U(1)$ symmetric spacetimes we are examining: We require asymptotic (in time) control of the conformal class of the two-dimensional spatial  geometry transverse to the $U(1)$ group orbits. In section \ref{se:VTD} below, we make this requirement (which we call the ``asymptotic conformal class condition") mathematically precise.

\section{VTD Equations and Solutions}
\label{se:VTD}

A spacetime has AVTD behavior if, as one approaches the singularity (chosen here to be at $t \rightarrow \infty$), the fields approach a solution of the VTD equations. One obtains the VTD equations from the Einstein equations by dropping all terms involving space derivatives, except in the momentum constraint (\ref{constr2}). Thus, if we use the notation $(\hat \gamma_{ab}, \hat \phi, \hat k_{ab}, \hat p)\leftarrow \rightarrow (\hat \sigma_{ab}, \hat \lambda, \hat \phi, \hat k_{ab},\hat  p)$ to denote these fields appearing in the VTD equations, we have the constraints 
\begin{equation}
\label{constr1VTD}
-\hat k^{ab} \hat k_{ab}+\hat \tau^2=2\hat p^2 
\end{equation}
and
\beq
\label{constr2VTD}
\hat \nabla_{b} \hat k_{a}^{b}-\partial _{a} \hat \tau =-2 \hat p \partial _{a} \hat \phi,
\eeq
and we have the evolution equations
\beq
\label{evolngammaVTD}
\partial_t \hat \gamma_{ab}= -2e^{\hat \lambda} \hat k_{ab}
\eeq 
and 
\beq
\label{evolnkVTD}
\partial_t \hat k_{ab} =-2\hat N \hat k_{ac} \hat k^c_b +\hat N \hat \tau\hat  k_{ab},
\eeq
together with
\beq
\label{evolnphiVTD}
\partial_t \hat \phi= e^ {\hat \lambda} \hat p.
\eeq
and
\beq
\label{evolnpVTD}
\partial_t \hat p= \frac{1}{2} \hat p \hat \psi  - \hat p \partial_t \hat \lambda.
\eeq
Note that to solve the VTD equations, we find it useful to rewrite (\ref{evolnkVTD}) in mixed index form:
\beq
\label{evolnkVTDmxd}
\partial_t \hat k_a^b = \hat N \hat \tau\hat  k_a^b.
\eeq

To enforce the asymptotic conformal class condition on these VTD solutions, we set the quantity $\hat \sigma_{ab}$ equal to a time-independent field $\tilde \sigma_{ab}$ on the 2 dimensional manifold $\Sigma^2$:
\beq
\hat \sigma_{ab}=\tilde \sigma_{ab}.
\eeq
(We shall generally use the tilde to denote time-independent quantities.)
It follows immediately from its definition that $\hat \psi=0$,  and it follows from (\ref{gammasigma}) and from (\ref{kdef}) that all of the time dependence in $\hat \gamma_{ab}$ and $\hat k_{ab}$ is contained in $\hat \lambda$. Consequently, from (\ref{evolnkVTDmxd}), we deduce 
\beq
\partial^2_{tt}\hat \lambda=0,
\eeq
which has the general solution
\beq
\hat \lambda =\tilde \lambda -\tilde v t,
\eeq
with $\tilde \lambda$ and $\tilde v$ a pair of arbitrary time-independent  functions on $\Sigma$. The general solutions for $ \hat \gamma_{ab}$ and $\hat k_{ab}$ now follow. We have, for example, $\hat k^a_b= \frac{1}{2}e^{-\hat \lambda}\delta^a_b \tilde v$

It remains to solve for $\hat \phi$ and $\hat p$. With $\hat \psi=0$, we may combine (\ref{evolnphi}) and (\ref{evolnpVTD}) to derive 
\beq
\partial^2_{tt} \hat \phi=0, 
\eeq
from which we obtain the general solution
\beq
\hat \phi=\tilde \phi -\tilde wt
\eeq
for the time independent functions $\tilde \phi$ and $\tilde w$. It follows that $\hat p=-e^{-\hat \lambda}\tilde w$.

What about the VTD constraint equations? If the solutions for $\hat k_{ab}$ and $\hat \tau$ and $\hat p$ are substituted into the VTD Hamiltonian constraint (\ref{constr1VTD}), then it reads
\beq
\label{HamconstrVTD}
e^{-2 \hat \lambda}(\frac{1}{2} \tilde v^2 -2\tilde w^2),
\eeq
which implies that $|\tilde w|=\frac{1}{2} |\tilde v|$. This reduces, up to signs of $\tilde v$ and $\tilde w$ the set of free functions in the VTD solution to  $\tilde v, \tilde \phi,  \tilde \lambda$,  and $\tilde \sigma_{ab}$. 

A similar substitution of the VTD solution quantities into the momentum constraint (\ref{constr2}) produces the equation
\beq
\label{momconstrVTD}
-\tilde v(\partial_b \tilde \lambda-\partial_b \tilde v t) + \partial_b \tilde v +4 \tilde w \partial_b (\tilde \phi-\tilde w t)=0
\eeq
(Here we need the $2+1$ identity $^3\Gamma^c_{ab} (\gamma) =\Gamma^c_{ab}(\sigma) +\frac{1}{2} (\delta ^c_b \partial_a \lambda + \delta ^c_a \partial_b \lambda -\sigma^{cd} \sigma_{ab} \partial_d \lambda).)$ We treat the case $\tilde w = \frac{1}{2} \tilde v$, then the momentum constraint can be written in the form 
\beq
\partial _b [\tilde v e^{(-\tilde \lambda +2 \tilde \phi)}]=0,
\eeq
which permits us t\textit{\textit{}}o solve for $\tilde v$, and thereby eliminate one more of the free functions in the asymptotic data; we are left with just $ \tilde \phi,  \tilde \lambda$ and $\tilde \sigma_{ab}$.

\section{AVTD Expansion  and the Fuchsian Analysis}
\label{se:VTD}

The procedure for using Fuchsian PDE methods to prove the existence of various families of solutions of Einstein's equations characterized by AVTD behavior near a spacelike singularity is by now fairly familiar \cite{KR98} \cite{IM02} \cite{DHRW02}.  One starts by writing out expressions for each of the fields as the sum of a solution of the VTD equations plus a remainder function multiplied by an exponential decay factor. This is done with no loss in generality, since the remainder fields are completely general. One then substitutes these expressions into the Einstein equations (first order form), and verifies that (perhaps with certain restrictions assumed on the free functions in the VTD solution) the remainder fields satisfy a PDE system in Fuchsian form \cite{KR98}. It follows that for each of the allowed solutions of the VTD equations, there is a  decaying solution for the remainder fields in the neighborhood of the singularity; consequently there is a solution of the Einstein equations which asymptotically approaches the given  solution of the VTD equations.

We find that this analysis is a bit simpler if we make a slight change in the choice of the field variables we use to describe our $U(1)$ symmetric spacetimes: we replace $\sigma_{ab}$ by its inverse $\sigma^{ab}$, and we replace $k_{ab}$ by $k_a^b$ (index raised by the metric inverse $\gamma^{ab}$). 
Then we write the expansion formulas for the fields as follows:
\beq
\label{expsigma}
\sigma^{ab} = \tilde \sigma^{ab} +e^{-\epsilon_{\sigma}t} \delta \sigma^{ab},
\eeq
\beq
\lambda=\tilde \lambda -\tilde v t +e^{-\epsilon_{\lambda}t} \delta \lambda,
\eeq
\beq
\phi=\tilde \phi -\tilde w t + e^{-\epsilon_{\phi}t} \delta \phi,
\eeq
\beq
\label{expk}
k_a^b=  e^{-\lambda} (\frac{1}{2} \tilde v  \delta_a^b + e^{-\epsilon_kt} \delta k_a^b),
\eeq
\beq
\label{expp} 
p= e^{-  \lambda}(-  \tilde w + e^{-\epsilon_pt} \delta p).
\eeq
Here $\epsilon_{\sigma}, \epsilon_{\lambda}, \epsilon_{\phi}, \epsilon_k$, and $ \epsilon_p$ are all positive numbers to be chosen later, while $\delta \sigma^{ab}, \delta \lambda, \delta \phi, \delta k_a^b$, and  $\delta p$ are the remainder fields. 

At this stage, it is useful to make a choice of gauge. Recalling that a choice of the lapse and shift largely control the gauge, and recalling (\ref{Nlambda}), we set $\delta \lambda=0$. That is, for a specified choice of the VTD data, we set
\beq
\label{explambda}
\lambda=\hat \lambda=\tilde \lambda -\tilde v t.
\eeq
For later purposes, we note that this gauge condition (\ref{explambda}) is equivalent to the condition
\beq
\label{gauge}
\frac{1}{2}\psi+ e^{-\epsilon_k t} \delta k^m_m=0.
\eeq
We recall that the shift has been set to zero. 

In checking whether the Einstein equations together with the above expressions for the initial data fields result in a Fuchsian PDE system for the remainder fields, we recall that to do this we must cast the system into first order form. Thus we need to add extra field variables, representing the spatial derivatives of those fields--$\phi$ and $\sigma^{ab}$-- whose second derivatives appear in the Einstein equations (\ref{constr1})-(\ref{evolnp}). To take care of the first of these, we introduce the new variable $\phi_{[c]}$, and a) everywhere first or second derivatives of $\phi$ appear in the Einstein evolution or constraint equations, we express  them in terms of $\phi_{[c]}$ and its first derivatives;  b) we assign to $\phi_{[c]}$ the evolution equation 
\beq
\label{phicevoln}
\partial_t \phi_{[c]} = \partial_c (e^{\lambda} p);
\eeq
and c) we assign to $\phi_{[c]}$ the expansion formula
\beq
\label{phicexpans}
\phi_{[c]}=\partial_c \tilde \phi - \partial _c\tilde wt + e^{-\epsilon_{\phi'} t} \delta \phi_{[c]}.
\eeq
Note that the expansion formula (\ref{phicexpans}) guarantees that if the remainder terms are bounded for both  $\partial_c \phi$ and $\phi_{[c]}$, then $\partial_c \phi$ and $\phi_{[c]}$ agree asymptotically.

Similarly, we also introduce the new variable $\sigma^{ab}_{[c]}$, and a) everywhere first or second derivatives  of $\sigma^{ab}$ appear in the Einstein evolution or constraint equations, we express them in terms of $\sigma^{ab}_{[c]}$ and covariant  derivatives $\tilde \nabla_m $ of $\sigma^{ab}_{[c]}$ (Here $\tilde \nabla_m $ denotes the Levi-Civita covariant derivative corresponding to the asymptotic metric $\tilde \sigma_{ab}$);  b) we assign to $\sigma^{ab}_{[c]}$ the evolution equation 
\beq
\label{sigmacevoln}
\partial_t \sigma^{ab}_{[c]}= \tilde \nabla_c (\partial_t \sigma^{ab})= (-\partial_c \tilde v) \sigma^{ab} - \tilde v \sigma^{ab}_{[c]} +2 e^{\lambda}[k^b_m \sigma^{ma}_{[c]} + \sigma^{am} \tilde \nabla_c k^b_m + (\partial_c \lambda) \sigma^{am} k^b_m],
\eeq
where we calculate $\partial_t \sigma^{ab}$ using $\sigma^{ab}=e^{\lambda} \gamma^{ab}$ , $\partial_t \gamma_{ab} =-2e^{\lambda} k_{ab}$,  and (\ref{explambda});
and c) we assign to $\sigma^{ab}_{[c]}$ the expansion formula
\beq
\label{sigmacexpans}
\sigma_{[c]}^{ab} = e^{-\epsilon_{\sigma'}t} \delta \sigma_{[c]}^{ab}.
\eeq
Again, these formulas are chosen to result in asymptotic agreement of $\sigma_{[c]}^{ab} $ and $\tilde \nabla_c \sigma^{ab}$, presuming that the remainder terms are bounded.

We are now ready to verify that the induced equations for the vector quantity
\beq
U:=(\delta \sigma^{ab}, \delta \phi, \delta k_a^b, \delta p,  \delta \sigma_{[c]}^{ab}, \delta \phi_{[c]})
\eeq
take the (first order) Fuchsian form
\beq
\label{Fuchs}
\partial_t U - L U = e^{-\mu t} F(x, t, U, \partial_a U),
\eeq
where $L$ is a linear operator which is independent of $t$ and has negative eigenvalues, $\mu$ is a positive number, and $F$ is  a function which is uniformly Lipschitzian in $U$ and $\nabla U$  so long as those quantities are uniformly bounded, and is analytic in $x$ and $t$. We do this verification for each of the components of $U$ separately, since as we shall see, the operator $L$ is diagonal. 

\subsection{Equations for $\delta \sigma^{ab}$ and $\delta \sigma_{[c]}^{ab}$}

To derive the evolution equation for $\delta \sigma^{ab}$ we work with the equation $\partial_t \gamma^{ab} =2e^{\lambda} k^{ab}$. For the right hand side, we use (\ref{expk}) and  (\ref{expsigma}) to calculate  $2e^{\lambda} k^{ab}=e^{-\lambda}(\tilde v  \sigma^{ab} +2e^{-2\epsilon_k t} \sigma^{ac}\delta k^b_c).$  For the left hand side, we use (\ref{explambda}) and  (\ref{expsigma}) to derive
$\partial_t \gamma^{ab} =\partial_t (e^{-\lambda} \sigma^{ab}) = e^{-\lambda}(\tilde v \ \sigma^{ab} + \partial_t \sigma^{ab})= e^{-\lambda}(\tilde v  \sigma^{ab} -\epsilon_{\sigma} e^{-\epsilon_{\sigma}t} \delta \sigma^{ab}+e^{-\epsilon_{\sigma}t} \partial_t \delta \sigma^{ab})$. Then combining, we obtain
\beq
\label{deltasigma}
\partial_t \delta \sigma^{ab} - \epsilon_{\sigma} \delta \sigma^{ab} = 2 e^{(\epsilon_{\sigma}-\epsilon_k)t}  
\sigma^{am}  \delta k_m^b
\eeq
This equation matches the requisite Fuchsian form so long as one chooses $\epsilon_k >\epsilon_{\sigma}>0.$ 

To obtain the evolution equation for $\delta\sigma_{[c]}^{ab}$ we substitute the various expansion formulas into (\ref{sigmacevoln}). We obtain, after some work,
\beq
\label{deltasigmac}
\partial_t \delta \sigma_{[c]}^{ab} -\epsilon_{\sigma'} \sigma_{[c]}^{ab} =2 e^{(\epsilon_{\sigma'}-\epsilon_k)t} \tilde \nabla_c( \sigma^{bm}\tilde  \delta k^a_m).
\eeq
We have Fuchsian form here so long as one chooses  $\epsilon_k >\epsilon_{\sigma'}>0.$ 

\subsection{Equation for $\delta k^b_a$}

Our derivation of the form for the evolution equation for $\delta k^b_a$ is based on the evolution equation for $k^b_a$, which is as follows
\beq
\label{evolnkdiag}
\partial_t k^b_a = N R^b_a +N \tau k^b_a - N \nabla ^b \nabla_a N -2N \gamma^{bm} \phi_{[m]}\phi_{[a]}.
\eeq
Using the expansion formulas, we now look at each of the terms in (\ref{evolnkdiag}):  We first calculate the time derivative term on the left hand side. From (\ref{expk}), we obtain
\beq 
\label{deltakderiv}
\partial_t k^b_a = e^{-\lambda}[ \frac{1}{2}\tilde v^2 \delta^b_a +(\tilde v-\epsilon_k) e^{-\epsilon_k t} \delta k^b_a + e^{-\epsilon_k t} \partial_t \delta k^b_a]. 
\eeq
We next calculate the $N \tau k$ term
\beq
N \tau k^b_a =  e^{-\lambda}[ \frac{1}{2}\tilde v^2 \delta^b_a +\tilde v e^{-\epsilon_k t} \delta k^b_a
+\frac{1}{2} e^{-\epsilon _k t}\tilde v \delta k^m_m \delta^b_a +e^{-2 \epsilon_kt} \delta k^m_m \delta k^b_a].
\eeq
Note that in this expression, as in (\ref{deltakderiv}), all but the first term has a factor $e^{-\epsilon _k t}$ (The first terms in these two expressions cancel). In the expressions we calculate below, no such factor appears.

For the hessian of N, we find
\beq
\nabla^b\partial_a N= e^{-\lambda} \sigma^{bm}\nabla_m \partial_a e^{\lambda} \\
= \sigma^{bm}[\partial_m \lambda \partial_a \lambda +\partial_m\partial_a \lambda -\Gamma^d_{am}(\gamma) \partial_d \lambda],
\eeq
where $\Gamma^d_{am}(\gamma)$ are the Christoffel symbols calculated from the metric $\gamma_{bc}. $ We could now expand this expression using the expansion formulas for $\sigma^{ab}, \sigma^{ab}_{[c]}, $ and $\lambda$.  However, the resulting formula is cumbersome, and all we really need to know is that it is 
analytic  in  $\delta \sigma^{ab}$ and $ \delta \sigma^{ab}_{[c]}, $ and that it has coefficients with at most 
polynomial growth in time. A similar analysis holds for the remaining two terms. For $NR^b_a$, since we are working on a two dimensional manifold, we have 
\beq
NR^b_a= e^{\lambda} R^b_a =\frac{1}{2} e^{\lambda} \delta^b_a R(\gamma)=\frac{1}{2} \delta^b_a [R(\sigma)- \Delta_{\sigma} \lambda],
\eeq
where $\Delta_{\sigma}$ is the laplacian corresponding to $\sigma$, and $R(\sigma)$ is its scalar curvature. This expression, if expanded out, is again analytic  in  $\delta \sigma^{ab}$ and $ \delta \sigma^{ab}_{[c]}$, and  $\tilde \nabla_d \delta \sigma^{ab}_{[c]} $ as well, with coefficients  growing no faster than polynomial in time. Finally for $-2N \gamma^{bm} \phi_{[m]}\phi_{[a]}$, we readily determine that this is analytic in $\delta \sigma^{mb}$ and $ \delta \phi_{[c]}$, with the same sort of coefficients.

Putting all of these expressions together, we have
\beq
\label{deltakeq}
\partial_t \delta k^b_a -\epsilon_k \delta k^b_a -\frac{1}{2}\tilde v \delta ^b_a \delta k^m_m= e^{-\epsilon_k t} \delta k^m_m \delta k^b_a +e^{(\epsilon_k -\tilde v)t} F^b_a(x, t, \delta \sigma^{ab}, \delta \sigma^{ab}_{[c]},  \tilde \nabla_d \delta \sigma^{ab}_{[c]},  \delta \phi, \delta \phi_{[c]}),
\eeq
where $F^b_a$ is a function of the indicated variables, and is polynomial in $\delta \sigma^{ab}, \delta \sigma^{ab}_{[c]},  \partial_d \delta \sigma^{ab}_{[c]},  \delta \phi$, and $\delta \phi_{[c]}$ with bounded coefficients depending on $x$ and $t$. Note that in deriving (\ref{deltakeq}), we have used the expansion $e^{(\lambda+\epsilon_k t)}=e^{(\epsilon_k-\tilde v )t} e^{\tilde \lambda}$.

To verify that this equation is of Fuchsian form, we find it useful to split it into its trace and trace-free pieces.  For the trace piece, we have 
\beq
\partial_t \delta k^b_b-( \epsilon_k +\tilde v)\delta k^b_b = e^{-\epsilon_k t} (\delta k^m_m)^2 +e^{(\epsilon_k -\tilde v)t} F^b_b(x, t, \delta \sigma^{ab}, \delta \sigma^{ab}_{[c]},  \partial_d \delta \sigma^{ab}_{[c]}, \delta \phi, \delta \phi_{[c]}).
\eeq
Clearly this is of Fuchsian form so long as $\tilde v(x)>\epsilon_k>0$. A similar observation holds for the trace-free piece.

\subsection{Equations for $\delta p$, $\delta \phi$ and $\delta \phi_{[c]}$}

We derive the evolution equation for the ``scalar field"  variable $\delta p$ (recall that $\phi, \phi_{[c]}$, and $p$ are really geometric variables) from the field equation (\ref{evolnp}) (with appropriate replacement of derivatives of $\phi$ and $\sigma^{ab}$ by $\phi_{[c]}$ and $\sigma^{ab}_{[c]})$ . To carry this out, we calculate each term in (\ref{evolnp}) using the expansion formulas for the relevant field variables, along with the imposed gauge condition (\ref{gauge}). 

We first calculate the term on the left hand side of equation  (\ref{evolnp}), $\partial_t p$:
\beq
\partial_t p=\partial_ t [e^{-\lambda}(-\tilde w +e^{-\epsilon_p t} \delta p)]
=e^{-\lambda}[\tilde v( -\tilde w+e^{-\epsilon_p t} \delta p) + e^{-\epsilon_pt}(-\epsilon_p \delta p+ \partial_t \delta p)].
\eeq
Next we obtain
\beq
-\frac{1}{2}p \psi - p \partial_t \lambda= e^{-\lambda} [-\tilde w + e^{-\epsilon_p t}\delta p][\tilde v + e^{-\epsilon_k t} \delta k^m_m].
\eeq
and
\beq
\label{laplacephiterm}
e^{\lambda} \gamma^{ab} \nabla_a\phi_{[b]}=\sigma^{ab} (\partial_a \phi_{[b]}-^3\Gamma^c_{ab} \phi_{[c]})= \sigma^{ab} \partial_a \phi_{[b]} -\sigma^{ab}  \Gamma^c_{ab} (\sigma) \phi_{[c]},
\eeq
where $^3\Gamma^c_{ab}$ denotes the $2+1$ dimensional spacetime Christoffel symbol, and $\Gamma^c_{ab}$ denotes its $2$ dimensional counterpart. These are related by the expression $^3\Gamma^c_{ab}=\Gamma^c_{ab} +\frac{1}{2} (\delta ^c_b \partial_a \lambda +\delta ^c_a \partial_b \lambda -\sigma^{cd}\sigma_{ab} \partial_d \lambda)$; however we note that in the calculation done in (\ref{laplacephiterm}), the $\lambda$ terms cancel. As with the hessian and curvature terms in the $\partial_t \delta k^b_a$ calculation, the full expansion for $e^{\lambda} \gamma^{ab} \nabla_a \phi_{[b]}$ is rather cumbersome. Again, it suffices to note that when we carry out this expansion, we obtain an expression which is analytic in $\delta \sigma^{ab}, \delta \sigma^{ab}_{[c]}, \delta \phi, \delta \phi_{[c]}$, and $\partial_d \delta \phi_{[c]}$,  with coefficients that have at most polynomial growth in time. 

Combining all of these terms, we obtain
\beq
\label{deltapeq}
\partial_t \delta p -\epsilon_p \delta p = -e^{(\epsilon_p-\epsilon_k)t} \tilde w \delta k^m_m +e^{-\epsilon_k t} \delta p \delta k^m_m + e^{(\epsilon_p-\tilde v)t}F(x,t, \delta \sigma^{ab}, \delta \sigma^{ab}_{[c]}, \delta \phi, \delta \phi_{[c]}, \partial_d \delta \phi_{[c]}, \delta p),
\eeq
where $F$ is a function of the type just described. We see that (\ref{deltapeq}) is in Fuchsian form so long as $\epsilon_k > \epsilon_p > 0$, and $\tilde v > \epsilon _p$.

The remaining evolution equations are very easily checked. From the equation $\partial_t \phi =e^{\lambda} p$, we obtain
\beq
\label{deltaphi}
\partial_t \delta \phi- \epsilon_\phi \delta \phi= e^{(\epsilon_\phi-\epsilon_p)t} \delta p,
\eeq
which is clearly of Fuchsian form so long as $\epsilon_p> \epsilon_\phi >0$.  Then from $\partial_t \phi_{[c]}= \partial_c (e^{\lambda}p)$ we derive
\beq
\label{deltaphic}
\partial_t \delta \phi_{[c]} -\epsilon _{\phi'} \delta \phi_{[c]} =e^{(\epsilon_{\phi'}-\epsilon_p)t} \partial_c \delta p,
\eeq
which is Fuchsian so long as $\epsilon_p>\epsilon_{\phi'} >0$.\\

\subsection{Applying the Fuchsian Theorem}

Taken together, the results of sections 4.1-4.3 show that, for any choice of the positive constants $\mathcal{E}:=(\epsilon_\sigma, \epsilon_k, \epsilon_{\sigma'}, \epsilon_{\phi}, \epsilon_p, \epsilon_{\phi'})$ and for any choice of the asymptotic data $\mathcal{A}:=( \tilde \sigma^{ab},  \tilde \phi, \tilde \lambda, \tilde v, \tilde w)$ satisfying the conditions 
\beq
\label{ineqs1}
\tilde v(x)>\epsilon_k> Max\{ \epsilon_{\sigma}, \epsilon_{\sigma'}, \epsilon_p\} 
\eeq
and
\beq
\label{ineqs2}
\epsilon_p >Max\{ \epsilon_\phi, \epsilon_{\phi'}\}
\eeq
the PDE system consisting of (\ref{deltasigma}), (\ref{deltasigmac}), (\ref{deltakeq}), (\ref{deltapeq}), (\ref{deltaphi}), and (\ref{deltaphic}) satisfied by the components of $U=(\delta \sigma^{ab}, \delta \phi, \delta k_a^b, \delta p,  \delta \sigma_{[c]}^{ab}, \delta \phi_{[a]})$ is of Fuchsian form. The following result is then a consequence of standard theorems regarding Fuchsian systems \cite{K96}

\begin{proposition}
For any choice of the positive constants $\mathcal{E}$ and analytical asymptotic data $\mathcal{A}$ which satisfy the conditions (\ref{ineqs1})-(\ref{ineqs2}), there is a unique analytic solution $U(t)$ of the equations (\ref{deltasigma}), (\ref{deltasigmac}), (\ref{deltakeq}), (\ref{deltapeq}), (\ref{deltaphi}), and (\ref{deltaphic}) for sufficiently large $t$. Further, this solution has the limit $\lim_{t\rightarrow \infty} U(t)=0$.
\end{proposition}

\section{AVTD Solutions of the Einstein Equations}
Since we have derived the Fuchsian system (\ref{deltasigma}), (\ref{deltasigmac}), (\ref{deltakeq}), (\ref{deltapeq}), (\ref{deltaphi}),  (\ref{deltaphic}) by substituting the expansion formulas for $( \sigma^{ab}, \sigma^{ab}_{[c]}, k^a_b, \lambda, \phi, \phi_{[c]}, p)$ into the Einstein evolution equations, we should be able to infer from Proposition 1 conclusions regarding solutions of Einstein's equations. To do this, we first need to verify that the unique solution $U$ whose existence is guaranteed by Proposition 1 satisfies $\partial_c \phi =\phi_{[c]}$ and $\tilde \nabla_c \sigma^{ab} =\sigma_{[c]}^{ab}.$ 

To make this verification, we use the evolution equations (\ref{sigmacevoln}) and (\ref{phicevoln})
to show that $\partial_t(\phi_{[c]}-\partial_c\phi)=0$ and $\partial _t(\sigma_{[c]}^{ab}-\tilde \nabla_c \sigma^{ab})=0$. Hence the quantities $\phi_{[c]}-\partial_c\phi$ and $\sigma_{[c]}^{ab}-\tilde \nabla_c \sigma^{ab}$ are independent of time. Next we  verify that $\lim_{t \rightarrow \infty}(\phi_{[c]} -\partial_c \phi)= \lim_{t \rightarrow \infty} ([\partial_c \tilde \phi -\partial_c \tilde w t +e^{-\epsilon_{\phi'} t }\delta \phi_{[c]}]-[\partial_c \tilde \phi -\partial_c \tilde w t +e^{-\epsilon_\phi t }\partial_c \delta \phi])= \lim_{t \rightarrow \infty} (e^{-\epsilon_{\phi'} t } \delta \phi- e^{-\epsilon_\phi t }\partial_c \delta \phi])=0.$ Similarly we calculate $\lim_{t\rightarrow \infty} (\sigma_{[c]}^{ab}- \tilde \nabla_c \sigma ^{ab}) =\lim_{t\rightarrow \infty} (e^{-\epsilon_{\sigma '}t} \delta \sigma^{ab}_{[c]}-e^{-\epsilon_{\sigma }t} \tilde \nabla_c  \delta \sigma^{ab})=0.$ It then follows that $\partial _c \phi(t)  =\phi_{[c]}(t)$ and $\tilde \nabla_c \sigma^{ab}(t)= \sigma^{ab}_{[c]}$ for all time $t$ during which the solutions exist, and so we have proven that we have a family--parametrized by the asymptotic data $\mathcal{A}$ --of solutions $(\gamma_{ab}(t), k_{cd}, \phi (t), p(t), N(t))$ of the Einstein \textit{evolution} equations for sufficiently large $t$, and further, the solutions are AVTD in a neighborhood of the singularity at $t \rightarrow \infty$.

It remains to show that, for certain choices of the asymptotic data $\mathcal{A}$, these solutions of the Einstein evolution equations  also satisfy the Einstein constraints, and hence the full Einstein system. We argue this via the following chain of observations and lemmas. 

We use the label $\mathcal{S}_{\mathcal{A}}(t)$ to denote the unique (AVTD) solution of the Einstein evolution equations which has asymptotic data $\mathcal{A}$, and we use the notation $C_\mu :=(C_0, C_a)= (-\frac{1}{2}[R-k^a_b k^b_a +\tau^2 -2p^2-2|\nabla \phi|^2], -[\nabla_bk^b_a-\nabla_a \tau+2p\nabla_a \phi])$ to denote the three constraint functions. We first note the following claim (which one proves by substituting the expansion forms of the dynamical variables into the constraint quantities $C_\mu$):

\begin{lemma}
For any choice of $\mathcal{A}$, the corresponding solution $\mathcal{S}_{\mathcal{A}}(t)$ satisfies the limit equation
\beq
\lim_{t\rightarrow \infty} C_\mu (\mathcal{S}_{\mathcal{A}}(t)) = \hat C_\mu (\mathcal{A}), 
\eeq
where $ \hat C_\mu$ denotes the VTD constraints (\ref{HamconstrVTD})-(\ref{momconstrVTD}). 
\end{lemma}

For a general choice of $ \mathcal{A}$, $\hat C_\mu (\mathcal{A})$ is nonzero. However, as shown in section 3, the VTD constraints are satisfied if and only if 
 $\tilde w=\frac{1}{2} \tilde v$, and $\tilde v=c e^{(\tilde \lambda- 2\tilde \phi)}$ for any constant $c$.  We denote by $\mathcal{A}_0$ those sets of data for which this is true.

We now seek a first order system of evolution equations for the constraint functions $C_{\mu}$. While one could derive these using the chain rule and the evolution equations for the dynamical variables, for present purposes we obtain the system in its most useful form by relying on the contracted Bianchi identity, which takes the form $D^{\nu} G_{\nu \mu}=0$, where $G_{\nu \mu}$ is the $2+1$ spacetime Einstein tensor. Combining this with the wave equation $D\cdot D\phi=0$ for $\phi$, we have 
\beq
\label{bianchid}
D_{\nu}\Theta^\nu_\mu=0,
\eeq
where $\Theta^\nu_\mu:=G^\mu_\nu-T^\nu_\mu$, so $\Theta^\nu_\mu=0$ corresponds to the Einstein equations. In particular, we have $\Theta^0_0=C_0$, and $\Theta^0_a=e^{-\lambda} C_a$ where $C_0$ and $C_a$ are the three constraint functions, written explicitly above. Expanding out (\ref{bianchid}), and using the $2+1$ decomposition expressions for the connection coefficients, together with some of evolution equations for the dynamical variables, we obtain
\beq
\partial_t C_0-2e^{\lambda} \tau C_0=\gamma^{ab}\nabla_a(e^{\lambda}C_b) + e^{\lambda}\gamma^{ab}(\nabla_a \lambda) C_b
\eeq
and
\beq
\partial_t C_a -e^{\lambda} \tau C_a=e^{\lambda} \nabla_a C_0 +2 e^{\lambda}(\nabla_a \lambda) C_0.
\eeq
Setting $E_0:=e^{2\lambda}C_0$ and $E_a :=e^{\lambda}C_a$, we may rewrite these equations in the form
\beq
\label{E0eq}
\partial_t E_0-2 \tilde v E_0 -2e^{\lambda}\tau E_0 =e^{\lambda}\sigma^{ab}\nabla_aE_b +\sigma^{ab} e^{\lambda}(\nabla_a \lambda)E_a
\eeq
and 
\beq
\label{Eaeq}
\partial_t E_a -\tilde v E_a-e^{\lambda}\tau E_a =\nabla_aE_0.
\eeq

We immediately note that, for any choice (solutions or not) of the functions $\lambda, \sigma^{ab}, \tau$ and $v$ which appear in the coefficients of (\ref{E0eq})-(\ref{Eaeq}), these equations are linear and homogeneous in $(E_0, E_a)$. Hence $(E_0, E_a)=(0,0)$ is always a solution. We want to show that for any data $\mathcal{A}_0$ for which the asymptotic condition $\hat C_\mu (\mathcal{A}_0)=0$ holds, $C_\mu (\mathcal{S}_{\mathcal{A}_0})=0$ necessarily holds. The key step towards that goal is to verify

\begin{lemma} If the coefficients for the system (\ref{E0eq})-(\ref{Eaeq}) are constructed from fields $\mathcal{S}_{\mathcal{A}_0}$ for which the asymptotic condition $\hat C_\mu (\mathcal{A}_0)=0$ holds, then the system (\ref{E0eq})-(\ref{Eaeq}) is Fuchsian
\end{lemma}
\textbf{Proof} Since the coefficients of these equations are constructed from solutions of the evolution equations, we may use the expansions and estimates derived in section 4.  Hence, we have
$e^{\lambda} \tau=\tilde v + e^{-\epsilon_k t} \delta k^m_m$, from which we determine that the left hand side of (\ref{E0eq}) is $\partial _t E_0 - 4 \tilde v E_0 - 2 e^{-\epsilon_k t} \delta k^m_m E_0$ and the left hand side of (\ref{Eaeq}) is $\partial_t E_a -2 \tilde v E_a -e^{-\epsilon_k t} \delta k^m_m E_a$. Removing the last term of each of these expressions (instead placing each on the right hand side of the respective equations), and recalling the inequality (\ref{ineqs1}), we see that the remaining expressions  $\partial _t E_0 - 4 \tilde v E_0$ and $\partial_t E_a -2 \tilde v E_a $ are clearly of proper Fuchsian form.
Recalling that $e^{\lambda} = e^{\tilde \lambda -\tilde v t}$, and noting the boundedness of $\sigma^{ab}$ and of $e^{-\epsilon_k t} \delta k^m_m$, we see that the right hand side of (\ref{E0eq}) together with the added term $4e^{-\epsilon_k t} \delta k^m_m E_0$ is of proper form as well. It remains to check the right hand side of (\ref{Eaeq}), together with its added term $2 e^{-\epsilon_k t} \delta k^m_m E_a$.

The added term is again clearly no trouble. As for the rest, writing out $E_0$ in terms of dynamical variables, we have
\beq 
\label{E00}
E_0= e^{2\lambda} C_0= -\frac{1}{2} e^{2 \lambda} [ R(\gamma) -k^a_b k^b_a +\tau^2 -2p^2 ]
\eeq
It follows from Proposition 1 that the quantity in the bracket in (\ref{E00}) is bounded for large $t$. Hence we can write
\beq
|\nabla_a E_0| \leq e^ {-\mu t} F(x,t),
\eeq
for some positive constant $\mu$ and some bounded function $F(x,t)$. This establishes that the right hand side, and therefore all of, (\ref{Eaeq}) is of Fuchsian form, so we have proven the Lemma.

Since the evolution equations for the constraint functions are (for appropriate asymptotic data) of Fuchsian form, it follows from the standard theorems that for such data, there is a unique solution which decays asymptotically. We already know, however, that $(C_0, C_a) =(0,0)$ is a solution. Therefore, $(C_0, C_a) =(0,0)$  is the only solution. We conclude that the constraints are satisfied by the fields $\mathcal{S}_{\mathcal{A}_0}$,  and we summarize our results as follows:

\begin{theorem} Let $\Sigma^2$ be any two dimensional manifold. For any set of asymptotic data $\mathcal{A}_0$ on $\Sigma^2$ which satisfies the asymptotic constraints and the asymptotic conformal class condition,  there is a unique polarized $U(1)$ symmetric solution of the vacuum Einstein equations on $S^1 \times \Sigma \times \mathbb{R}$ which has AVTD behavior in the neighborhood of its singularity. 
\end{theorem}

\section{Conclusions} 

The existence of a large family of $U(1)$ symmetric solutions with AVTD behavior has been established in earlier work  \cite{IM02}. The significance of this work is that we now know that for any two dimensional manifold $\Sigma$, there is such a family of AVTD solutions on  $\Sigma \times S^1 \times \mathbb{R}$. 

As noted in the Introduction, the goal of this work is to be able to produce solutions with AVTD behavior in a neighborhood of the singularity, and with global (proper time) existence evolving away from the singularity. One possible way to do this is to show that there is an AVTD solution of the sort discussed here, which admits a constant mean curvature (in the two dimensional sense) surface for which an appropriate energy for the data on these surfaces satisfies a smallness condition. It is not yet clear whether or not this can be done.
\\

\section{Acknowledgments}
We thank the Kavli Institute for 
Theoretical Physics at Santa Barbara, the National Center for Theoretical Sciences in Hsinchu, and L'Institut des Hautes \'Etudes Scientifiques at Bures-sur-Yvette for providing very pleasant and stimulating environments for our collaboration on this work. This work was partially supported by the NSF, under grants PHY-0099373 and PHY-0354659 at Oregon,  and grants PHY-0098084 and PHY-0354391at Yale.

\end{document}